\documentstyle[psfig]{mn}

\title{An iterative technique for solving equations of statistical equilibrium}

\author[L. B. Lucy]
       {L. B. Lucy \\
        Astrophysics Group, Blackett Laboratory, Imperial College of Science,
        Technology and Medicine, Prince Consort Road, London SW7 2BW\\
E-mail: l.lucy@ic.ac.uk}

\begin{document}

\maketitle

\begin{abstract}
Superlevel partitioning is combined with a simple relaxation
procedure
to construct an iterative technique for solving equations of statistical
equilibrium.
In treating an $N$-level model atom, the technique avoids the $N^{3}$ scaling
in computer time for direct solutions with standard linear
equation
routines and also does not fail at large $N$ due to the accumulation of
round-off errors. In consequence, the technique allows detailed model atoms
with $N \ga 10^{3}$, such as those required for iron peak elements, to be  
incorporated into diagnostic codes for analysing astronomical spectra.
Tests are reported for a 394-level Fe II ion and a 1266-level Ni I--IV atom.
\end{abstract}


\section{Introduction}
In order to extract all the information contained in an
astronomical
spectrum, the structure of the emitting layers must be modelled and the
resulting spectrum
computed. In doing this, various technical challenges must be overcome,
including in particular that
of solving the equations of statistical equilibrium for appropriately 
detailed atomic models of the relevant elements. 
Now, in some cases, quite simple atomic models suffice,
as when predicting the strengths of collisionally-excited forbidden lines in
the spectra of gaseous nebulae
(e.g., de Robertis et al. 1987). But when treating
line formation in stellar atmospheres and winds or in novae and supernovae
ejecta, the important contibutions of the iron group elements Sc--Ni require
atomic
models with numbers of energy levels $N \ga 10^{3}$ per atom (e.g.,
Hubeny \& Lanz 1995). The direct
solution of the equations of statistical equilibrium with a standard linear
equation-solving package is then decidedly problematic. As Press et al. (1992, p.23)
note, solving systems with $N$ as large as several hundred is feasible with
double precision arithmetic but beyond this the limiting factor is generally
machine time. This limitation arises because the required computing
time scales as $N^{3}$ (e.g., Press et al. 1992, p.34).

	Classically, this computational problem was completely 
circumvented by simply assuming that the levels are in local thermodynamic 
equilibrium (LTE); and essentially
the same strategy can in be used to approximate reality more closely
if the
Boltzmann--Saha LTE formulae are replaced by formulae that take account of the
dominant NLTE effect in the problem of interest. This strategy  has been
followed in Monte Carlo codes for stellar winds (Abbott \& Lucy 1985;
Lucy \& Abbott 1993) and 
supernovae (Lucy 1987; Mazzali \& Lucy 1993; Lucy 1999) where the dominant
NLTE effect is dilution of the radiation field due to spherical extension.  
But this approach, despite its theoretical basis, has been criticized
by Hillier \& Miller (1998) as    
giving an
ionization structure and line source functions that are {\em ad hoc} or, at best,
inconsistent with the radiation field. On the other hand, Springmann \& Puls
(1998) find an encouraging degree of agreement between the ionization formula
and accurate NLTE calculations.         

	The solution strategy preferred by Hillier \& Miller (1998)  and 
indeed by most other investigators 
is that pioneered by Anderson (1989) in
which a reduced atomic model is constructed by consolidating neighbouring
energy levels into single superlevels. In this way, $N$ is reduced to
$\la 50$, so that solution with a standard package is possible even with
single precision arithmetic (Press et al. 1992, p.23). Then, from the
the reduced atom's level populations, those for the complete
atom -- needed
for the transfer calculation -- are obtained by assuming that all the
sublevels
of a given superlevel have the same departure coefficient.  

	In the earliest implementations of Anderson's seminal idea, the
partitioning into superlevels was based solely on excitation energy (Anderson
1989, Dreizler \& Werner 1993). But subsequently Hubeny \& Lanz (1995)
introduced the constraint that all sublevels of a given superlevel must be of
the same parity. This eliminates the awkward problem of radiative transitions
between a superlevel's sublevels as well as enhancing the validity of the
assumption of identical departure coefficients. Nevertheless, for some
diagnostically-important emission lines, Hillier \& Miller (1998) still find
it necessary to solve for the populations of the individual emitting 
levels.

	Evidently, current treatments of the statistical equilibria of complex
atoms represent interim procedures pending more powerful computers or
better solution strategies. An example of the latter is the subject of this
paper.

	The technique described and tested herein arose in an ongoing effort
to extend the range and sophistication of problems treatable with Monte Carlo
methods
(Lucy 1999a,b). But since the technique itself is more widely
applicable, it is presented here separated from its Monte Carlo origins.

\section{An iterative scheme}

In this section, an iterative scheme is described for determining the 
statistical equilibrium of a model atom with large N in a specified 
environment
-- i.e., the ambient radiation field is assumed known as are the electron
density and temperature. Because of these assumptions, this scheme
will often be used within an outer iteration loop determining the
thermal structure of the emitting layers as well as the internal radiation
field.

\subsection{Basic idea}

In the problems of primary interest, the ambient radiation field,
although not black body, nevertheless derives from thermal processes. For
such a non-pathological radiation field, the dominant ions of a given species
have heavily
populated low levels, especially the ground state and low-lying metastable
levels, and sparsely populated high levels. In this circumstance, the
populations of high levels are largely determined by their coupling to
the low levels of the same ion via bound-bound (b-b) transitions and to the
low levels of the upper ion via free-bound (f-b) transitions. Accordingly,
to a first approximation, the population of a high level is independent of
the populations of other high levels of the same or of other ions. This then
suggests that the populations of high levels be estimated one by one with a
simple relaxation scheme that brings such levels into statistical equilibrium
with the readily estimated populations of low levels. Then, having thus
derived approximate populations for high levels, we can obtain improved
populations for the
low levels with the same relaxation scheme.

	The simple, fast iterative technique just described often yields
accurate level populations for a single ion with far less computer time than
required by the direct solution of the basic equations (Li, McCray \&
Sunyaev 1993). But in some
circumstances, intial convergence is followed by divergence,
and so the technique lacks the robustness needed if it is to be included
in a code of wide applicability. Accordingly, this relaxation procedure is
here incorporated into the superlevel technique discussed in 
Section 1. By combining these procedures, the
desired robustness is achieved without sacrificing speed. Moreover, in
contrast to the superlevel technique, there is no limit to the achievable
accuracy.
 
\subsection{Equations of statistical equilibrium}

Level $i$ of the complete model atom is in statistical equilibrium if
\begin{equation}
 R_{\ell i}n_{\ell}-(R_{i \ell}+R_{iu})n_{i}+R_{ui}n_{u} = 0
\end{equation}
where a summation convention is adopted for upper ($u>i$) and lower
($\ell<i$) levels. The quantity $R_{ij}$ denotes the total rate coefficient
for the transition $i \rightarrow j$, and $n_{i}$ denotes the relative
population of level $i$ -- i.e., we impose the normalization constraint
\begin{equation}
 \sum_{i} n_{i} = 1
\end{equation}
	This normalization constraint together with statistical equilibrium 
equations for $N-1$ levels constitute a system of $N$ equations in the $N$
unknowns $n_{i}$. If the ambient conditions are given, the coefficients
$R_{ij}$ are independent of $n_{i}$ and the system is linear.

	Note that equation (1) applies to all levels of all ions included in
the complete atomic model. Accordingly, $R_{ij}$ refers to b-b transitions
when $i$ and $j$ denote levels of the same ion and to b-f or f-b transitions
when they denote levels of different ions. 

\subsection{Superlevels}

Levels consolidated into a superlevel will be referred to as sublevels
of the parent superlevel, and the complete set of superlevels defines the
reduced atomic model. Rate coefficients coupling superlevels determine the
statistical equilibrium of the reduced atom exactly as do the $R_{ij}$ for
the complete atom -- see equation (1). Moreover, these rate coefficients must
be defined   
so that, if the populations of their sublevels are known exactly,
the statistical equilibrium of the reduced atom yields the exact populations
for the
superlevels. Clearly, this is a necessary condition if the iterative
scheme is to converge to the exact solution for the complete atom.

\subsubsection{Coupling of superlevels}

Let $j = j_{1},j_{2},...$ be an index denoting the sublevels of
superlevel $J$ and let $n_{j}$ be the {\em exact} populations of these
sublevels. The population of the superlevel is then naturally defined to be
\begin{equation}
  n_{J}= \sum_{j} n_{j}
\end{equation}
from which it follows that
\begin{equation}
 \sum_{J} n_{J} = 1
\end{equation}

	Now consider transitions between superlevels $J$ and $K$. The rate
$R_{JK}n_{J}$ at which the hypothetical transition $J \rightarrow K$ transfers population
from superlevel $J$ to superlevel $K$ must equal (for the exact solution) the
sum of the rates $R_{jk}n_{j}$ at which the actual transitions $j \rightarrow k$  
transfer population from the sublevels of $J$ to the sublevels of $K$.
Thus
\begin{equation}
  R_{JK} n_{J} = \sum_{j} \sum_{k} R_{jk} n_{j}
\end{equation}
and correspondingly for the inverse coefficient $R_{KJ}$.

	With rate coefficients for superlevels thus defined, the equations of
statistical equilibrium for the complete model atom, when summed over the
sublevels of individual superlevels, yield the corresponding equilibrium
equations for the reduced atom. Accordingly, if $n_{i}$ is the exact solution
vector for equations (1) and (2), then the corresponding superlevel vector $n_{J}$
from equation (4) is the exact solution for the reduced atom.

\subsubsection{Positivity} 

A complete set of approximate rate coefficients $R_{JK}$ for the
reduced atom
can be computed from equation (5) if the vector of population {\em ratios}
 $\, r_{j} = n_{j}/n_{J}$ is estimated for each superlevel $J$. Provided these
estimates $r_{j}$ are $>0$ for all $j$, the coefficients $R_{JK}$ will be
$>0$, and so the equations of statistical equilibrium for the reduced atom,
having exactly the form of equations (1) and (2), will give $n_{J}>0$ for
all J.  

\subsection{Iteration steps} 

The iterative scheme for solving the equations of statistical
equilibrium for the complete atomic model now proceeds step by step as
follows:

        1) From the current estimate for the level populations $n_{i}$ of
the complete atom, the population ratios $r_{j} = n_{j}/n_{J}$ for each
superlevel's
sublevels are computed and used in equation (5) to derive rate
coefficients for transitions between superlevels.  

        2) The equations of statistical equilibrium for the reduced atomic
model are solved directly with a standard linear
equation-solving package, thus obtaining improved populations
$n_{J}$ for the superlevels.

        3) The populations of each superlevel's sublevels are rescaled by
setting $n_{j} = n_{J} r_{j}$, where $n_{J}$ are the updated values from 
step 2) and $r_{j}$ the ratios from step 1). With this rescaling, the current
estimate $n_{i}$ for the complete
atom now satisfies the normalization constraint.

        4) The populations of each ion's energy
levels are adjusteded sequentially from highest to lowest
according to the following relaxation scheme     
\begin{equation}
 n_{i} = \frac{R_{\ell i}n_{\ell} +R_{ui}n_{u}}{R_{i \ell}+R_{iu}}
\end{equation}
where, as in equation (1), summation over the indices $\ell$ and $u$ is
implied.

	5) The vector $n_{i}$ is renormalized.

	6) The current estimate $n_{i}$ is accepted if the adopted
convergence criterion is satisfied.

        7) If the criterion is not satisfied, a further iteration is
initiated by inputting $n_{i}$ at step 1).

\subsection{Comments} 
 
The following comments are intended to clarify the working of the iterative 
scheme:

        a) To start the iterations, an initial estimate for the level
populations
is required. An obvious choice is the Saha--Boltzmann LTE population, and this
is how the numerical examples of Section 3 are initiated. But when the
scheme is implemented within an outer iteration loop for the thermal
structure,
better initial estimates are available after the first outer iteration.

        b) From the initial estimate, only the ratios
$r_{j} = n_{j}/n_{J}$ are actually used. Thus
the scheme may be regarded as iterating on these ratios. 
 
        c) Equation (6) gives the value of $n_{i}$ that satisfies 
equation (1) for level $i$, given the current estimates of the
vectors $n_{\ell}$ and $n_{u}$.

        d) The value of $n_{i}$ from equation (6) is
used in $n_{u}$ when  $n_{i-1}$ is calculated. Thus, improved values are
used immediately rather than waiting for the next iteration cycle.

        e) If we estimate the level populations of an ion, we are typically 
fairly accurate for the well-populated low levels and progressively less so
for higher levels. This consideration is behind the {\em downward} sweep of
the relaxation step. The least accurately determined levels are then
adjusted first, and so their improved populations are available for the
summations 
$R_{ui} n_{u}$
when later adjusting the lower levels of the same ion.

        f) Step 4) clearly preserves the positivity of $n_{i}$. Together with
the remarks in Section 2.3.2, this proves that the iterative scheme as a
whole always returns $n_{i}>0$ for all $i$ provided only that their initial
values are $>0$. 

	g) Transitions between sublevels of the same parent superlevel do not
appear in equation (5) and so make no direct contribution to the solution
vector $n_{J}$ obtained at step 2). But all such transitions appear in 
equation (6). Accordingly, after the first iteration, these transitions 
contribute indirectly to $n_{J}$ as the relaxation step adjusts the ratios
$r_{j}$ and hence also the rate coefficients $R_{JK}$.  

	h) In view of the physical meaning of both the numerator and
denominator in equation (6), no problems with loss of accuracy due to the
accumulation of roundoff errors will arise from the relaxation step even 
if the
scheme is applied to atomic models with extremely large $N$. Accordingly,
such problems
should not arise for the iterative scheme provided the reduced atom has an
appropriately limited $N$.

	i) If the scheme converges, the particular choice of partitioning into
superlevels becomes irrelevant. In contrast, with the standard superlevel
technique, the partioning determines and limits the accuracy attained.

	j) Steps 4) and 5) may be repeated several times before proceeding to
step 6).

\subsection{Convergence}

If the iterative scheme is initiated with (or reaches) the exact solution
$n_{i}^{(x)}$ of equations (1) and (2), then, as
discussed in Section 2.3.1, steps 1) and
2) return the exact populations of the superlevels. 
Step 3) then recreates $n_{i}^{(x)}$, and this is left unchanged by
step 4), since every level is found to be already in statistical equilibrium.
On the other hand, any vector $n_{i} \neq n_{i}^{(x)}$, and which therefore
does not satisfy equations (1) and (2), will be changed at step 4).
Accordingly, the 
iterative scheme converges {\em if and only if} it reaches the exact
statistical
equilibrium solution for the {\em complete} atomic model.

	Note that if the rate coefficient $R_{ij}$ is changed
for the transition between any pair of sublevels belonging to the same
superlevel, the solution for the complete atom
changes but that for the reduced atom may not -- cf. comment g) of
Section 2.5. Accordingly, there may exist vectors $n_{i} \neq n_{i}^{(x)}$
that
are unchanged by steps 1) to 3). Equation (6) is therefore fundamental
to the `if and only if' assertion in the above convergence
statement.

	For practical applications, the rate of convergence to $n_{i}^{(x)}$
is of prime concern. Unfortunately, no general quantitative results have
been established on this question. But in the limiting case where each
superlevel comprises just one sublevel, the iterative scheme reduces to the
direct solution of equations (1) and (2) and so converges at the first
iteration.
As we increasingly depart from this limit by consolidating more and more
levels into fewer and fewer superlevels, the rate of convergence will surely
slow and, eventually, the scheme might even diverge. Fortunately, numerical
experience
indicates that, with the degree of consolidation typical of current
implementations of the superlevel technique, the scheme does converge.
Moreover, the rate of convergence is such
that solutions of the desired accuracy are obtained with a huge reduction
of computer time relative to that required by direct solution.   

\section{Numerical experiments}

In this section, the iterative scheme described in Section 2 is
applied to model atoms having the completeness of level structure needed
for modern spectral synthesis calculations. No particular attempt has been
made to
compile the best modern data for these atoms since the intent
is merely to illustrate the typical performance of the scheme.  

\subsection{Convergence criteria}

In practical applications of iterative schemes, the exact solution is
not known and so convergence is often judged from the
corrections to the previous iterate. For atoms with large $N$, it is
convenient to summarize the corrections to all the levels in a single
quantity. Here we weight all levels equally and compute the mean of the
absolute fractional changes to the individual levels' populations.
Thus the adopted measure of the correction at the $r$th iteration is 
\begin{equation}
 \delta^{(r)}=\frac{1}{N} \sum_{i} \frac{| n_{i}^{(r)} - n_{i}^{(r-1)} |}{n_{i}^{(r-1)}}
\end{equation}
	By now, a well-recognized danger in judging convergence on the basis
of small iterative corrections is that an acceptance criterion of the form
$ \delta^{(r)} < \delta$ does not necessarily imply that the typical
fractional error is $\delta$. The remaining errors will in fact be much
larger if
the scheme converges slowly. Accordingly, in the following numerical
experiments, for which the exact solutions $n_{i}^{(x)}$ are known, the
quantity $\delta^{(r)}$ is compared to
\begin{equation}
 \epsilon^{(r)}=\frac{1}{N} \sum_{i} \frac{| n_{i}^{(r)} - n_{i}^{(x)} |}{
 n_{i}^{(x)}}
\end{equation}
In this way, we may hope to establish values of $\delta$ required to achieve
a specified absolute accuracy.

\subsection{Fe II}

As a first test, the scheme is applied to the 394-level model Fe II ion used
previously (Lucy 1999b) to investigate treatments of line emissivity in
a Monte Carlo spectral synthesis code for supernovae. Indeed, it was these
emissivity experiments that led to the relaxation scheme that forms step 4)
of the present technique.

	The energy levels of the Fe II ion and the f-values for permitted
transitions were extracted from the Kurucz--Bell (1995) compilation by
M.Lennon (Munich). Einstein A-values for forbidden transitions are from
Quinet et al.
(1996) and Nussbaumer \& Storey (1988). Collision strengths are from
Zhang \& Pradhan (1995), with the van Regemorter (1962) formula as default
for permitted transtions and the value 0.05 as default for forbidden
transitions.

	As previously (Lucy 1999b), the ambient radiation field is taken to be
$W B_{\nu}(T_{b})$ with $T_{b} = 12500K$ and dilution factor $W = 0.067$
corresponding to the point $r=2R$ above the photosphere. But now instead of
the low density limit, electron excitation is included with $T_{e} = 10000K$
and $N_{e} = 10^{7} cm^{-3}$.

	In order to solve for the Fe II ion's level populations in these
ambient conditions with the iteration scheme of Section 2, the 394 levels
must be partitioned into superlevels. Because of the importance of the lowest
levels, the 16 levels of the terms $a\: ^{6}\!D$, $a\:^{4}\!F$,
$a\:^{4}\!D$ and $a\:^{4}\!P$
are treated individually. Thereafter, taking account of the physical
arguments of Hubeny \& Lanz (1995), levels are consolidated into a single
superlevel if they arise from terms having the same multiplicity and parity
and are not widely separated in energy. The resulting reduced ion has 43
levels.   
 
\subsubsection{Iterations}

        Figure 1 shows the performance of the iterative scheme for the
specified ambient conditions and with the chosen partitioning. The initial
estimate -- the zeroth iterate -- is the Boltzmann population at $T_{e}$
whose accuracy relative to the exact solution $n_{i}^{(x)}$ is plotted at
$r=0$. Thereafter, success in converging to $n_{i}^{(x)}$ is measured by
$\epsilon^{(r)}$ and the iterative corrections to $n_{i}$ by $\delta^{(r)}$.
From the plotted values of $\epsilon^{(r)}$, we see that there is a rapid
initial improvement followed by slow but steady subsequent improvements,
with $\epsilon^{(r)}$ then decreasing by about 0.1 dex per iteration.

\begin{figure}
\vspace{5.5cm}
\includegraphics{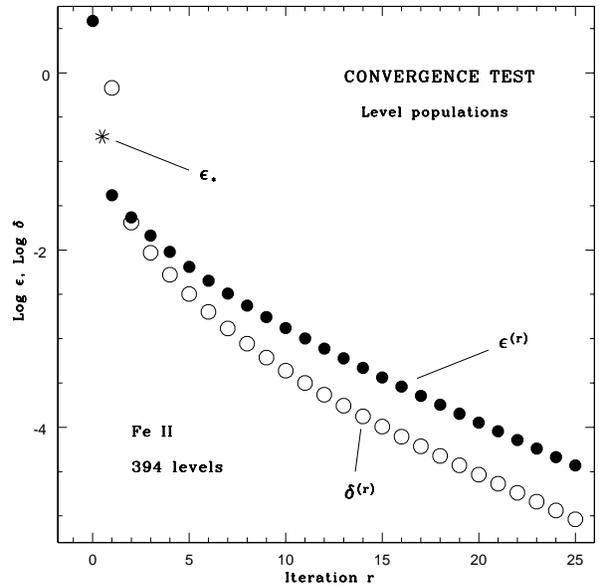}
\caption{Convergence test for 394-level model Fe II ion. The quantities
$\delta^{(r)}$ and
$\epsilon^{(r)}$ defined in the text are plotted against iteration number
$r$. The point labelled $ \epsilon_{*}$ and plotted at $r=1/2$ is the value
of $\epsilon$ for the solution obtained with the superlevel technique.  }
\end{figure}

        Because of this relatively slow convergence, a large number of
iterations are needed if very high accuracy is demanded. Thus, while only 4
iterations give $\epsilon < 10^{-2}$, 12 are required to achieve 
$\epsilon < 10^{-3}$, 21 for $\epsilon < 10^{-4}$ and 32 for   
$\epsilon < 10^{-5}$. But given the parametric and structural uncertainties
of astrophysical
models and the limited accuracy of atomic data, there is no practical
need for high accuracy at present. Solutions with $\epsilon \la 10^{-2}$ 
are entirely adequate for current diagnostic investigations of
observed spectra.

        The quantity $\epsilon^{(r)}$ measures the
typical accuracy achieved for the level populations. But the largest
fractional error among all 394 levels is
also recorded at each iteration. For
the test plotted in Figure 1, $ \epsilon_{max} < 10^{-1}$ at $r=3$,
$ < 10^{-2}$ at $r=11$ and $ < 10^{-3}$ at $r=19$. 

        In practical applications, the exact solution is not known and so
convergence must be decided using a quantity such as $\delta^{(r)}$. Figure 1
shows that, after a few iterations, $\delta^{(r)}$ is significantly smaller
than $\epsilon^{(r)}$, a situation familiar from iteration schemes for
radiative transfer problems. In fact, for large $r$, we find that
$\delta^{(r)} / \epsilon^{(r)} \approx 0.25$. Accordingly, this test 
suggests that a convergence criterion of the form  
$\delta^{(r)} < \gamma \epsilon$ with $\gamma \approx 0.25$
will give a solution for which the typical fractional error of $n_{i}$
levels is $\epsilon$. Unfortunately, the speed of convergence and the
relationship between 
$\delta^{(r)}$ and $\epsilon^{(r)}$ are problem-specific and undoubtedly also
partitioning-specific. This particular test illustrates the relatively
difficult
case near a photosphere where significant radiative excitation of high
levels occurs. On the other hand, in very dilute, UV-strong radiation
fields, where the
normal levels are severely depopulated, the convergence is rapid.

	To illustrate this point, the above test is repeated with parameters
$T_{b} = 40000K$,
$W = 7.9 \times 10^{-13}$, $T_{e} = 10000K$ and $N_{e} = 4000 cm^{-3}$,
representing conditions at a typical point in the Orion Nebula (Lucy 1995).
In this case, the scheme achieves $\epsilon^{(r)} < 10^{-3}$ already at
$r = 5$, at which
point $\epsilon^{(r)}$ is decreasing by about 0.64 dex per iteration. Notice
that this extremely rapid convergence occurs despite the very poor initial
estimate, namely
the Boltzmann population at $T_{e}$.

\subsubsection{Asymptotic behaviour}

The nearly constant logarithmic decrements in $\delta^{(r)}$ and
$\epsilon^{(r)}$ for large $r$ seen in Figure 1 indicate that the iterative
scheme has a simple
asymptotic behaviour, and this becomes strikingly evident when the plot is
extended to $r \ga 40$, for then the two sets of points become linear and
parallel. Although no rigorous results have been established,
inspection of residuals suggests that
\begin{equation}
 \frac{n_{i}^{(r)} - n_{i}^{(x)}}{n_{i}^{(r-1)} - n_{i}^{(x)}}\;\: \rightarrow
\;\:\alpha \;\;\;\;as \;\;\;\; r \; \rightarrow \;\infty
\end{equation}
with the constant $\alpha \approx 0.81$ for this Fe II test problem.

	A useful application of this numerically-suggested asymptotic
behaviour is to
estimate
$\epsilon^{(r)}$ even when the exact solution $n_{i}^{(x)}$ is not known.
It readily follows from equation (9) that as $r \rightarrow \infty$
\begin{equation}
\frac{\delta^{(r)}}{\delta^{(r-1)}} \sim \alpha \;\;\;\; and \;\;\;\;
\epsilon^{(r)} \sim  \frac{\alpha}{1-\alpha}\: \delta^{(r)}
\end{equation}
Together these formulae allow $\epsilon^{(r)}$ to be estimated for $r \geq 2$.
Comparison of these estimates with the exact values of  $\epsilon^{(r)}$
reveals acceptable accuracy for $r \geq 4$ and the
expected high accuracy for large $r$. At $r=4$, the estimate is
$ 0.72 \times $ the exact value.  

	Another possible application 
is to predict $n_{i}^{(x)}$ from consecutive iterates
$n_{i}^{(r-1)}$ and $n_{i}^{(r)}$ and thereby achieve accelerated
convergence.
Thus, after the $r$th iteration, we would replace $n_{i}^{(r)}$ by the
estimate  
\begin{equation}
 n_{i}^{(x)}=n_{i}^{(r)} + \frac{\alpha}{1-\alpha} \:
[\:n_{i}^{(r)} - n_{i}^{(r-1)}\:] 
\end{equation}
and then continue the iterations at step 1). But this modification is not
in fact recommended since it
jeopardizes the scheme's robustness and, in any case, by the time the 
asymptotic behaviour is established,
satisfactory accuracy has already been achieved. Nevertheless,
if and when highly accurate solutions are required for atoms with large $N$,
this device for accelerating convergence could be exploited, perhaps
stabilized with an undercorrection factor. 

\subsubsection{Computer time}

This iterative scheme has been designed so that, when treating model atoms
with $N \ga 10^{3}$, the full $N^{3}$ scaling of direct solutions is avoided.
Clearly, $N^{3}$ scaling still applies for step 2), but with
$N$ here reduced from 394 to 43, the time required by the direct solution for
the reduced atom
is reduced dramatically -- by a factor $ \approx 1.3 \times 10^{-3}$ --
compared to
that required for the complete atom.
This leaves the initial calculation of the coefficients $R_{ij}$ and,
during each iteration, the calculation of the superlevel coefficients $R_{JK}$
and the relaxation step, all of which scale as $N^{2}$, as the major users
of CPU.

	The actual times required on a Sun Ultra 1 with clock speed 170 MHz
and with 256 Mb of memory are as follows. The
direct solution for the 394-level ion takes 21.15s, of which 0.70s is the
time required to calculate the $R_{ij}$.

	In contrast, the time required for the iterative scheme to obtain a 
solution with 1 percent accuracy is 1.30s. This comprises the 0.70s for the
$R_{ij}$ and 3 iterations, each of which takes 0.20s. Within an iteration, 
the breakdown is
0.13s for the  $R_{JK}$, 0.02s for the direct statistical equilibrium 
solution for the reduced ion, and 0.05s for the relaxation step.

	Although not providing acceptable accuracy, the time required by
the superlevel technique is 0.85s. This comprises the times required for
calculating the 
$R_{ij}$ and $R_{JK}$ coefficients in addition to the trivial time required
for the direct solution of the reduced ion.        

	Timing experiments have also been carried out to test the efficacy
of performing $m$ relaxation and normalization steps per iteration cycle --
see comment j) of Section 2.5. The Fe II test was repeated with $m = 1-4$,
and the results compared by plotting the accuracy parameter $\epsilon^{(r)}$
not against $r$ but against $t^{(r)}$, the elapsed computer time. This
comparison shows that the scheme with $m = 2 \;or\; 3$ is initially marginally
superior but that
for $r \ga 16$ the standard scheme with $m = 1$ is best. 

\subsection{Ni I--IV}

In view of the scheme's success in treating a detailed model Fe II
ion,
we now test its ability to treat ionization as well as excitation by 
applying it to a multi-ion model of Ni, an atom closely comparable to Fe with
regard to the complexity
of its level structure. 

	The model Ni atom has 1266 levels, comprising a 186-level model for
Ni I, 717 levels for Ni II, 344 for Ni III, and a severely truncated 19-level
model for Ni IV. These energy levels and the f-values for permitted
transitions between them were extracted from the Kurucz--Bell (1995)
compilation. Data for forbidden transtions are from Garstang (1964) for Ni I,
from Nussbaumer \& Storey (1982) for Ni II, and from Garstang (1958) and
Osterbrock (1992) for Ni III. The van Regemorter (1962) formula is used to
compute collision strengths for permitted transitions and 1.0 is
the default value for forbidden transitions. Also included are
photoionizations to and radiative recombinations from the well-populated  
lower levels of the ions Ni II--IV. The photoionization cross sections are 
assumed to follow Seaton's formula with $\beta = 2$ and $s = 2$ (Pauldrach et
al. 1994) and to have hydrogenic threshold cross sections. The corresponding
total radiative recombination coefficients are derived from Milne's relation.
Collisional ionization is neglected.  

	The partitioning of the Ni ions into superlevels follows the procedure
adopted for Fe II, with several of the lowest levels left single and higher
levels grouped into superlevels when they have the same multiplicity and
parity and are not widely separated in energy. Specifically, the Ni I levels
are consolidated into 24 superlevels with the first 7 being single, the
corresponding numbers are 33 and 8 for Ni II, 27 and 9 for Ni III, and 2 and 0
for the truncated Ni IV ion. The net result of this partitioning is to reduce
a model atom with 1266 levels to one with 86 superlevels. 

	As for the Fe II test problem, the ambient radiation field is taken
to be $W B_{\nu}(T_{b})$ but now with $T_{b} = 7500K$ and dilution factor
$W = 0.5$. This value of $W$ corresponds to a point at the photosphere 
($r = R$), and
thus the importance of radiative excitation of high levels is enhanced 
relative to the Fe II test. Electron collisional excitation is included
with $T_{e} = 6500K$ and $N_{e} = 10^{8} cm^{-3}$.

\subsubsection{Iterations}

Because of the huge value of $N$, no attempt
has been made to derive the exact solution $n_{i}^{(x)}$ with a linear
equation-
solving package. Instead, $n_{i}^{(x)}$ is taken to be the solution
obtained after 50 iterations of the iterative scheme, and this is used 
when calculating $\epsilon^{(r)}$ from equation (8)  

	Figure 2, drawn to the same scale as Figure 1, shows the performance
of the iterative scheme for the specified ambient conditions and with the
chosen partitioning of the Ni atom. The initial
estimate is the Saha--Boltzmann population at $T_{e}$.
From the plotted values of $\epsilon^{(r)}$, we see that there is again a
rapid
initial improvement followed by slow but steady subsequent improvements,
with $\epsilon^{(r)}$ then decreasing by about 0.07 dex per iteration, a
somewhat slower rate of convergence for large $r$ than that seen in Figure 1.

\begin{figure}
\vspace{5.5cm}
\includegraphics{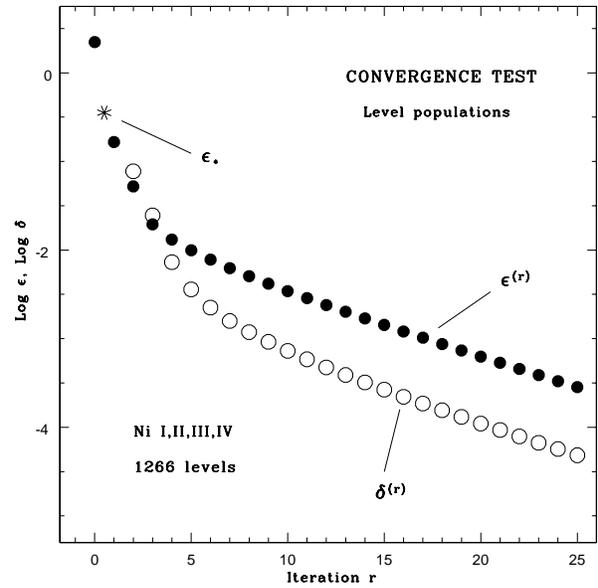}
\caption{Convergence test for 1266-level model Ni atom. The quantities
$ \delta^{(r)}$ and
$ \epsilon^{(r)}$ defined in the text are plotted against iteration number
$r$. The point labelled $ \epsilon_{*}$ and plotted at $r=1/2$ is the value
of $\epsilon$ for the solution obtained with the superlevel technique.  }
\end{figure}

	As in the Fe II test, the largest 
fractional error among all 1266 levels of the complete atom is 
also recorded at each iteration. For
the test plotted in Figure 2, $ \epsilon_{max} < 10^{-1}$ at $r=6$,
$ < 10^{-2}$ at $r=18$ and $ < 10^{-3}$ at $r=32$. 
	
	The nearly constant logarithmic decrements in $\delta^{(r)}$ and
$\epsilon^{(r)}$ for large $r$ evident in Figure 2 indicate that the
iterative scheme's simple
asymptotic behaviour when applied to a single ion also applies for a
multi-ion atomic model.
Inspection of residuals confirms that equation (10) again holds
asymptotically, but with
slightly different coefficients $\alpha$ for the different ions.

\subsubsection{Error distributions}

In Figure 3, the distribution of the absolute values of the 
logarithmic errors in the level
populations
obtained with the superlevel technique -- i.e., after step 3) of the
first iteration -- is compared with the corresponding error
distributions after the first three iterations of the
iterative scheme.

\begin{figure}
\vspace{7.8cm}
\includegraphics{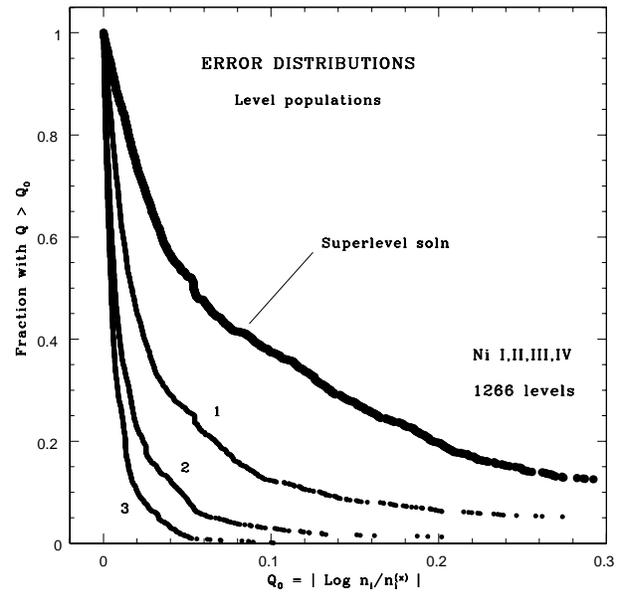}
\caption{Distributions of the absolute values of the logarithmic errors
for the estimated level populations
for the 1266-level model Ni atom. The error distributions are plotted for the 
solution given by the superlevel technique as well as for the first three
iterates obtained with the iterative scheme.}
\end{figure}

	Now, given the accuracy expected of modern diagnostic
investigations of astronomical spectra, errors in level populations of
0.1 dex are surely the maximum acceptable. But for the superlevel solution
plotted
in Figure 3, no less than 37.4
percent of the levels have errors exceeding 0.1 dex, and there is an extended 
high-error tail to beyond 0.3 dex. Thus, despite a physically-sensible
partitioning and
a relatively large $N$ for the reduced atom, the superlevel solution, if used
in a diagnostic analysis, would significantly limit the accuracy of that 
analysis. This conclusion is in accord with the finding of Hillier and Miller
(1998) that the superlevel technique is not sufficiently accurate for
diagnostically-important emission lines.    

	The comparison in Figure 3 of the errors of the superlevel solution
with those after 1 iteration of the iterative scheme shows the dramatic
effect of the relaxation step in immediately decreasing the number of levels
with substantial errors. Levels with errors exceeding 0.1 dex drop to 12.2
percent in consequence of the relaxation step, and then drop to 2.9
and 0.2 percent with two further iterations. Clearly, in this case, even with
as few as 3
iterations, the iterative scheme provides a solution that could
be used in a diagnostic analysis with the confident expectation that errors
in solving for statistical equilibrium are unlikely to be compromising the
accuracy of the results.  

\subsubsection{Computer time}

By scaling from the Fe II case, we can estimate that the direct 
solution for the 1266-level would take 678s, assuming that the build up of
round off errors did not in fact prevent a solution. This is to be compared
with
an iteration time of 4.66s. The breakdown of this 
is 2.08s to calculate the $R_{JK}$, 0.18s for the direct solution of the 
86-level reduced atom, and 1.40s for the relaxation step. With just a few
iterations needed to achieve the desired accuracy, there is evidently a
vast saving in computer time. Moreover, the direct solution is in any case
not a realistic option.

\subsection{Line formation}

In the above test problems, the ambient radiation field is a featureless
continuum. But typically $J_{\nu}$ will be strongly affected by line
formation, and so it is important to investigate how this impacts on the
performance of the scheme.

        The simplest test incorporating line formation assumes that the point
considered is in an expanding flow and treats line formation in the Sobolev
approximation. The b-b rate coefficients are then modified as follows
(Klein \& Castor 1978): line trapping is included by replacing the Einstein
coefficient $A_{ul}$ by $A_{ul}\beta_{ul}$, where $\beta_{ul}$ is the
Sobolev escape probability; and the profile-averaged mean intensity, needed
for the rates of radiative excitation and de-excitation, is
$\beta_{ul}J_{lu}$, where $J_{lu}$ is the mean intensity of the   
unattenuated continuum

	With these modifications, the rate coefficients $R_{ij}$ depend on the
Sobolev optical depth and therefore on $n_{i}$. In consequence, the
statistical equilibrium problem is no longer linear. Fortunately, simple
repeated back substitutions (Lucy 1999b) allow  $n_{i}^{(x)}$ to be
determined with high accuracy in $\sim 5$ iterations.  

	A second consequence of the dependence of the $R_{ij}$ on $n_{i}$
is that, in applying the iterative scheme, the $R_{ij}$ must be recalculated
at the beginning of each iteration cycle before calculating the coefficients
$R_{JK}$ for the superlevels.

\subsubsection{Test problem}

The effect of line formation
on the iterative scheme is now investigated by repeating the test of
Section 3.2,
but we now suppose that the point considered is in a supernova's 
homologously expanding envelope,
that the elapsed
time since explosion is 13 days, and that the number density of Fe II ions is 
$5.1 \times 10^{5} cm^{-3}$. These two additional parameters determine the
Sobolev
optical depths of the b-b transitions and are such that the converged solution
has numerous optically-thick UV lines, with optical depths up to 
$2 \times 10^{4}$.

	As with the previous tests, convergence is monitored by computing
$\epsilon^{(r)}$, and this reveals {\em faster} convergence than when line
formation is ignored, with $\epsilon^{(r)}$ now dropping by 0.15 dex per
iteration at large $r$ as against 0.1 dex previously. Because of this faster
convergence, only 2 iterations
are required to achieve $\epsilon < 10^{-2}$, 4 for $\epsilon < 10^{-3}$,
9 for $\epsilon < 10^{-4}$ and 16 for $\epsilon < 10^{-5}$.

	As in the more extreme Orion Nebula case of Section 3.2.1, the
faster convergence in this test can be attributed to the reduced populations
of high normal levels. When high levels are depleted, either by dilution or
by line formation, the validity of the assumption motivating the relaxation
step  -- that
high levels are negligibly coupled to other high levels -- is
strengthened -- see Section 2.1. 

\section{Conclusion}

The aim of this paper has been to describe and test an iterative technique 
for obtaining approximate solutions to the equations of statistical
equilibrium with the
accuracy required by modern diagnostic analyses of astronomical spectra. In
particular, the technique is designed to treat model atoms whose numbers of
levels $N$ are so large that exact solutions with standard software
packages are precluded, either
because of excessive computer time or because of an anticipated loss of 
accuracy due to the
accumulation of round-off errors. By combining Anderson's (1989) superlevel
technique with a simple relaxation method, solutions of the desired accuracy
are derived in $\sim$ 3--5 iterations with a huge saving of computer time compared
to direct solutions, and for values of $N$ for which direct solutions are not
a feasible option.    

	As mentioned in Section 1, this technique was developed for Monte
Carlo codes. But the most promising immediate application would seem to be
to the already-existing Accelerated Lambda Iteration (ALI) codes that use 
superlevel partitioning to avoid treating atomic models with excessively
large $N$. In view of Figure 3 and the earlier work of Hillier \& Miller
(1998), the accuracy of these codes almost certainly falls short of
minimum requirements for the quantitative analyses of observed spectra.
Fortunately, the iterative technique described herein would seem to
be readily incorporated into ALI codes and would bring them to the desired
accuracy with little impact on their complexity, robustness or use of
computer time.

\end{document}